\documentclass[conference]{IEEEtran}
\IEEEoverridecommandlockouts

\usepackage{array} 
\usepackage{gensymb}
\usepackage{makecell}
\usepackage{multirow}
\usepackage{bbding}
\usepackage{cite}
\usepackage{amsmath,amssymb,amsfonts}
\usepackage{algorithmic}
\usepackage{graphicx}
\usepackage{textcomp}
\usepackage{xcolor}
\def\BibTeX{{\rm B\kern-.05em{\sc i\kern-.025em b}\kern-.08em
    T\kern-.1667em\lower.7ex\hbox{E}\kern-.125emX}}
\begin{document}

\title{Location-Oriented Sound Event Localization and Detection with Spatial Mapping and Regression Localization\\

\author{
    \IEEEauthorblockN{Xueping Zhang$^{1}$, $^{*}$Yaxiong Chen$^{2,1,3}$, Ruilin Yao$^{1}$, Yunfei Zi$^{1}$, $^{*}$Shengwu Xiong$^{2,1,3}$\thanks{* Corresponding Author}}    
    \IEEEauthorblockA{$^1$ School of Computer Science and Artificial Intelligence, Wuhan University of Technology, Wuhan 430070, China}
    \IEEEauthorblockA{$^2$ Sanya Science and Education Innovation Park, Wuhan University of Technology, Sanya 572000, China}
    \IEEEauthorblockA{$^3$ Shanghai Artificial Intelligence Laboratory, Shanghai, 200232, China}
    \IEEEauthorblockA{\{xpzhang, chenyaxiong, yaoruilin, yfzi, xiongsw\}@whut.edu.cn}
}

\thanks{This work was supported in part by the National Key Research and Development Program of China (Grant No. 2022ZD0160604) and NSFC (Grant No. 62176194), in part by Hainan Province “Nanhai New Star” Technology Innovation Talent Platform Project under Grant NHXXRCXM202361, in part by the Youth Fund Project of Hainan Natural Science Foundation under Grant 622QN344.}
}

\maketitle

\begin{abstract}
Sound Event Localization and Detection (SELD) combines the Sound Event Detection (SED) with the corresponding Direction Of Arrival (DOA). Recently, adopted event-oriented multi-track methods affect the generality in polyphonic environments due to the limitation of the number of tracks. To enhance the generality in polyphonic environments, we propose Spatial Mapping and Regression Localization for SELD (SMRL-SELD). SMRL-SELD segments the 3D spatial space, mapping it to a 2D plane, and a new regression localization loss is proposed to help the results converge toward the location of the corresponding event. SMRL-SELD is location-oriented, allowing the model to learn event features based on orientation. Thus, the method enables the model to process polyphonic sounds regardless of the number of overlapping events. We conducted experiments on STARSS23 and STARSS22 datasets and our proposed SMRL-SELD outperforms the existing SELD methods in overall evaluation and polyphony environments.
\end{abstract}


\section{Introduction}
\begin{figure}[htbp]
\centerline{\includegraphics[width=0.5\textwidth]{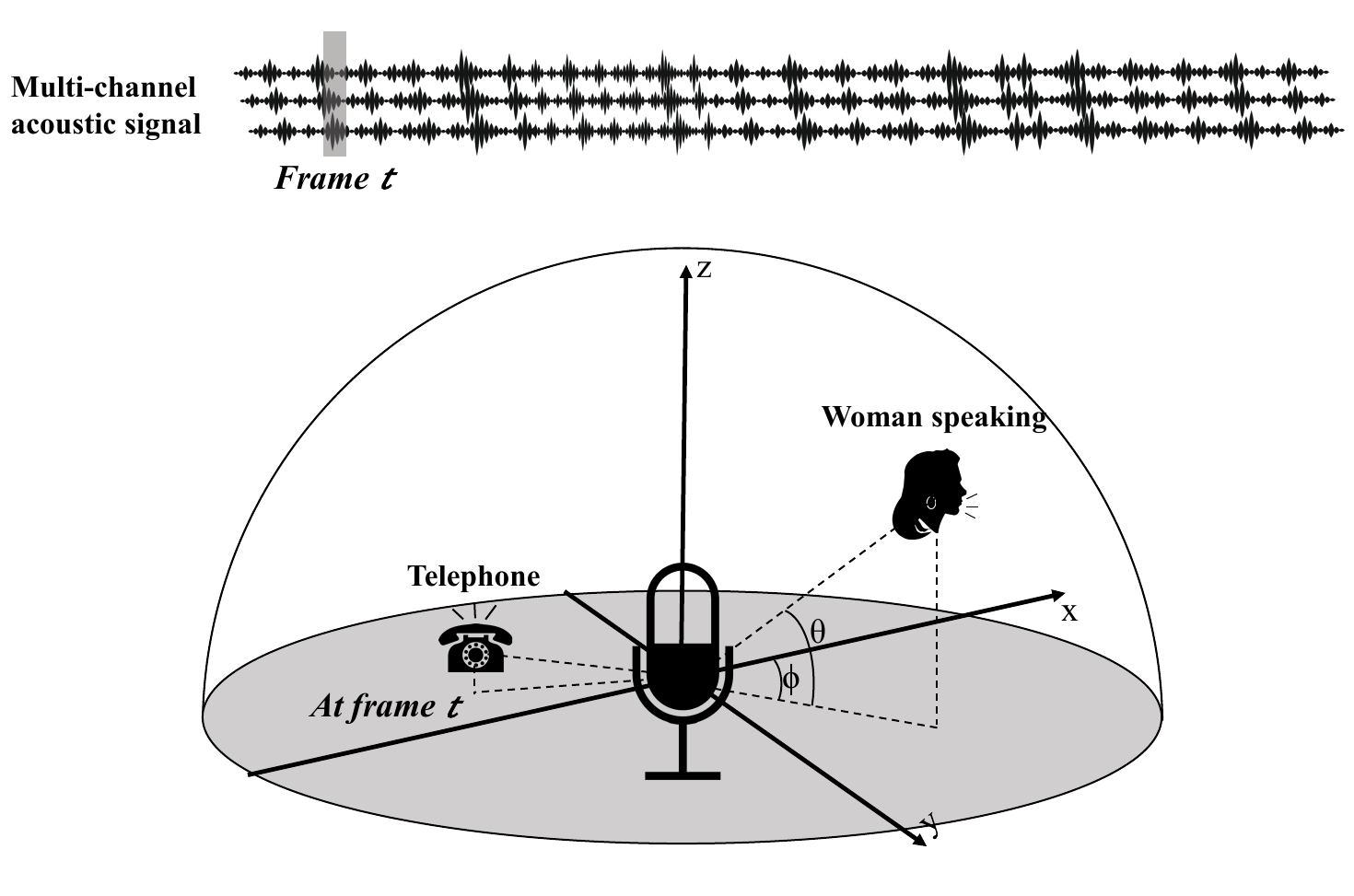}}
\caption{The class and location of events occurring at frame $t$ in a multi-channel acoustic signal in a 3D space. The location is described by azimuth $\phi$ and elevation $\theta$.}
\label{SELD}
\end{figure}

Sound Event Localization and Detection (SELD) \cite{b_seld1, b_seld, b_seld3} is a technology that uses multi-channel acoustic signal processing technology to identify the class of sound events in audio and determine the time and spatial location of the sound. As shown in Fig.~\ref{SELD}, SELD has two subtasks: Sound Event Detection (SED) \cite{b_sed1, b_sed2, b_sed3, b_sed4} and identification of the Direction-Of-Arrival (DOA)\cite{b_doa1, b_doa2, b_doa3}. SED determines the class of sound events (e.g., Telephone, Woman speaking) at frame $t$ in a multi-channel acoustic signal. DOA determines the direction of the sound source in a 3D space at frame $t$, often described by azimuth $\phi \in [-180\degree, 180\degree]$ and elevation $\theta \in [-90\degree, 90\degree]$. SELD has played an essential role in many applications, such as surveillance \cite{b_surveillance1, b_surveillance2}, bio-diversity monitoring \cite{b_bio}, and context-aware devices \cite{b_context}. In recent years, with the development of deep learning, SELD research has made significant progress.

The prevalent SELD methods can be categorized into three output formats. 
The first is the class-wise output format \cite{b_single1, b_ACCDOA}, in which the model predicts activities of all event classes and corresponding locations. Adavanne et al.\cite{b_single1} proposed SELDnet, which detects sound events and estimates the corresponding DOAs using two branches: a sound event detection (SED) branch and a DOA branch. Activity-coupled Cartesian DOA (ACCDOA) \cite{b_ACCDOA} assigns an event activity to the length of a corresponding Cartesian DOA vector, which enables a SELD task to be solved without branching. 
The second is the track-wise output format\cite{b_cao, b_sequence}, that is, each track detects an event and the corresponding location. Cao et al. \cite{b_cao} proposed an event-independent network (EINV2) and incorporated Permutation Invariant Training (PIT) \cite{b_pit} into a SELD task to solve a track permutation problem. However, the track-wise output method cannot cope with the situation where events of the same class occur at different locations.
Although the above two output formats have achieved some success, they are unable to cope with the situation where events of the same class occur at different locations. Therefore, the track-class output format\cite{b_mutil_ACCDOA1} was proposed. This method expands the class-wise vector output method to track-wise. It can handle situations where events of the same class occur at different locations, thus making up for the shortcomings of the multi-track output format.

Moreover, researchers \cite{b_Improving, b_lavss} apply the sound separation to SELD to separate the overlapping sound events from different locations, and Wang et al.\cite{b_single3} employed spatial augmentation techniques to deal with data sparsity in SELD, broadening the scope of potential solutions.

\begin{figure*}[htbp]
\centerline{\includegraphics[width=1\textwidth]{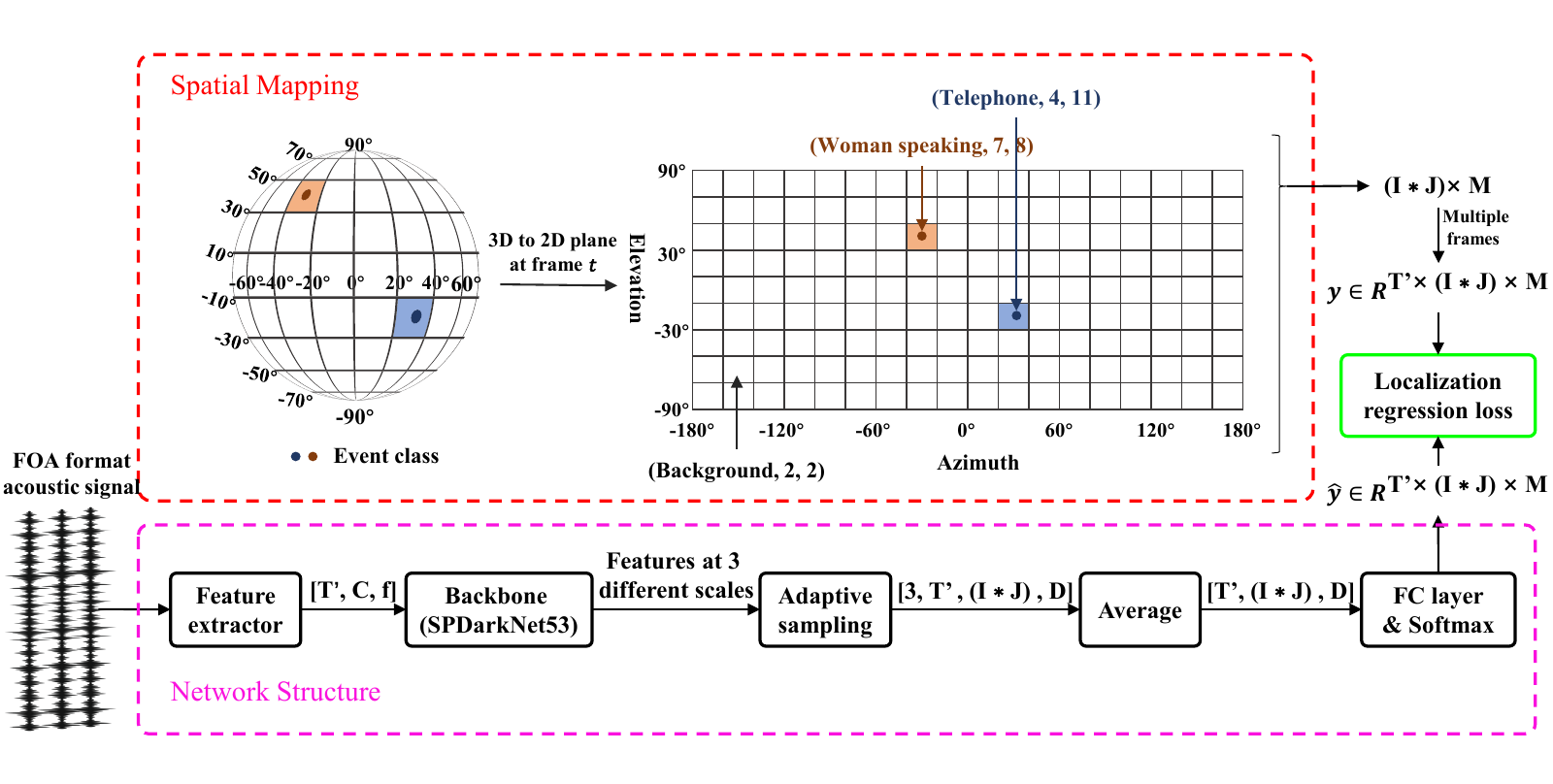}}
\caption{Schematic of our location-oriented sound event localization and detection method, including three parts: Spatial Mapping, Network Structure, and Localization regression loss. $[\cdot , \cdot , ...]$ represents shape of the features.}
\label{main}
\end{figure*}

While the aforementioned research has seen success in SELD, both the aforementioned methods need prior knowledge of the maximum polyphony count when solving the polyphonic problem. However, in real-world applications or with unlabeled data, this number is often unknown. This uncertainty limits the generality of these methods. 
To overcome the challenge of unknown maximum polyphony, in this paper, we shift away from the traditional event-oriented approach and propose a location-oriented method, namely Spatial Mapping and Regression Localization for SELD (SMRL-SELD). Specifically, we segment the 3D spatial space, mapping it to a 2D plane, and propose a regression loss to guide the localization. This method can predict the events at every location without being constrained by the number of overlapping events or the knowledge of the maximum polyphony. Our contributions are as follows:
\begin{enumerate}
\item We propose a location-oriented method, SMRL-SELD, which segments the 3D spatial space into a 2D plane, giving a new solution for polyphonic problems.
\item We propose a regression localization loss to guide localization, enhancing the ability to accurately detect and localize sound events in a polyphonic environment.
\item All experiments are conducted on the STARSS23\cite{b_STARSS23} and STARSS22\cite{b_STARSS22} datasets. The results show that SMRL-SELD outperforms existing SELD methods in both the overall evaluation and polyphonic environments.

\end{enumerate}

\section{Method}
As shown in Fig.~\ref{main}, our location-oriented SELD method first transforms labels via spatial mapping and then obtains predictions using a multi-scale neural network. The model is then optimized via a localization regression loss function. This section will provide a detailed introduction to the three components: spatial mapping, network structure, and localization regression loss function.

\subsection{Spatial Mapping}

The first-order Ambisonics (FOA) format is obtained by converting the 32-channel microphone array signals by means of encoding filters based on anechoic measurements of the Eigenmike array response. 
The FOA signal consists of four channels $(W, X, Y, Z)$ with $W$ corresponding to an omnidirectional microphone and $(X, Y, Z)$ corresponding to three bidirectional microphones aligned on the Cartesian axes. 

We are given a FOA format sound source corresponding to $T$ frames. Within each frame $t$, there are $ N $ active events, where $ N \geq 0 $. For the $n\text{-}th$ reference target among the $ N $ events at frame $ t $, it can be defined by the triplet $ \{c_n, \phi_n, \theta_n\} $. Here, $c_n$ represents the class of the sound event, and the polar coordinate $( \phi_n, \theta_n )$ is the position of the sound event at frame $t$.

The polar coordinate $( \phi_n, \theta_n )$ can represent any position in a 3D plane. First, we segment the 3D plane into a grid consisting of $I$ rows and $J$ columns. Then we unfold the 3D plane into a 2D plane, as shown in Fig.~\ref{main}. The azimuth $\phi_n$ and elevation $\theta_n $ are represented by the indexes of the corresponding grid cells. In this way, any polar coordinate system coordinates can be mapped into the 2D plane with corresponding grid cells.
Then, for the $n\text{-}th$ reference target among the $N$ events at frame $t$, it can be redefined on the 2D plane by the triplet $\{c_n, i_n, j_n\}$. Here, the coordinate $(i_n, j_n )$ refers to the grid cell at the $i$-th row and $j$-th column on the 2D plane. 

In addition to the classes defined in the dataset (e.g., ‘Telephone’, ‘Woman speaking’, etc.), we have also included the background sound as a new class ‘Background’. In this way, we can give each grid cell in the 2D plane a reference target label. For the whole 2D plane, there are $(I \ast J)$ labels, where $I$ and $J$ represent the number of rows and columns in the grid respectively. We convert these labels to one-hot format. For $T'$ frames of the model input, we get the 2D format reference target labels $ y \in R^{T’ \times (I \ast J)  \times M}$, where $M$ represents the number of all event classes in the dataset including the ‘Background’ and `$\ast$' represent the multiplication.

\subsection{Network Structure}
We use a feature extractor to process FOA format acoustic signal to get the multi-channel acoustic feature $x \in R^{T' \times C \times f}$, where $C$, $T'$ and $f$ denote the number of channels, time frames, and dimensions of the input acoustic features, respectively.
We use CSPDarkNet53 in yolov8 \cite{b_yolov8} as the backbone network. The backbone network converts the multi-channel acoustic feature $x$ into the representation $\hat{x}$. 
Benefit from the multi-scale expression ability of the backbone, the hidden state $\hat{x}$ has three different scales, which we adaptively resample to the shape of $[3, T’, (I \ast J), D]$ to align them along the spatial dimension, where $I$ and $J$ represent the number of rows and columns of the grid respectively, depending on the size of grid partitioning (grid size), $D$ is the size of the hidden layer for the class dimension.
Then we average the multi-scale hidden state $\hat{x}$ along the scale dimension to obtain the embedding $x^{\prime} \in R^{T’ \times (I \ast J)\times D}$.
Then we utilize a fully connected layer and softmax operation along the class dimension to get the predicted probability distribution $ \hat{y} \in R^{T’ \times (I \ast J)  \times M}$, across $(I \ast J)$ grid cells and $ T’$ frames, where $M$ denotes the number of all event classes in the dataset including the ‘Background’.

\subsection{Localization Regression Loss}
After obtaining 2D format reference target labels and predicted probability distributions, we calculate a localization regression loss. The core idea of this loss function is that we regard the localization and detection on the 2D plane as a task like object detection\cite{b_object1, b_object2}.
When a 3D sphere is transformed into a 2D plane, the detection task encounters challenges similar to those in object detection, such as significant class imbalance. To address these issues, a localization regression loss is proposed for our model training. This regression loss is composed of three components: 1) a class-wise mean square error loss function, 2) an area intersection union ratio loss function, and 3) a converging localization loss function.

\subsubsection{Class-wise Mean Square Error Loss Function}
The class-wise mean square error loss function $L_{Class\text{-}MSE}$ directly reflects the difference between the predicted event and the reference labels, the formula is shown in \eqref{mse}, where $I$ and $J$ represent the number of rows and columns of the grid on 2D plane, respectively. The $\hat{y}_{gm}$ and $y_{gm}$ represent the predicted and reference labels, respectively.
\begin{eqnarray}
\begin{array}{l}
L_{Class\text{-}MSE}=\frac{1}{(I \ast J) \times M} \sum_{g=1}^{(I \ast J)}\sum_{m=1}^{M}\left(y_{gm}-\hat{y}_{gm}\right)^{2}
\end{array}
\label{mse}
\end{eqnarray}

\subsubsection{Area Intersection Union Ratio Loss Function}
We define events that are not `background', such as `Telephone' and `woman talking', as `non-background'. The area intersection union ratio loss function $L_{AIUR}$ calculates the ratio of intersection and union between the predicted non-background areas and reference target in the 2D plane, as shown in \eqref{iou}. It reflects the degree of overlap between the predicted non-background areas and the reference target and measures the model's localization ability. The closer the ratio is to 1, the more precise the localization is. 
$y_{g} \in \{0, 1\}$ represents takes the value of 0 for background and 1 for non-background, and $\hat{y}_{g} \in \{R \mid 0<R<1\}$ denotes the predicted probability of non-background events in the corresponding DOA of the $g\text{-}$th grid cell. The symbols `×' and `+' denote the multiplication and addition of the corresponding elements, respectively.
\begin{eqnarray}
\begin{array}{l}
L_{AIUR}=1-\frac{\sum_{g=1}^{G}\left(y_{g}\times\hat{y}_{g}\right)}{\sum_{g=1}^{G}\left(y_{g} + \hat{y}_{g} - y_{g}\times\hat{y}_{g}\right)} ,  \\ \\
y_{g}= \begin{cases}
  0 ,&   \text{if  background}  \\
  1 ,&   \text{if  non-background} 
\end{cases}
\end{array}
\label{iou}
\end{eqnarray}

\begin{figure}[t]
\centerline{\includegraphics[width=0.45\textwidth]{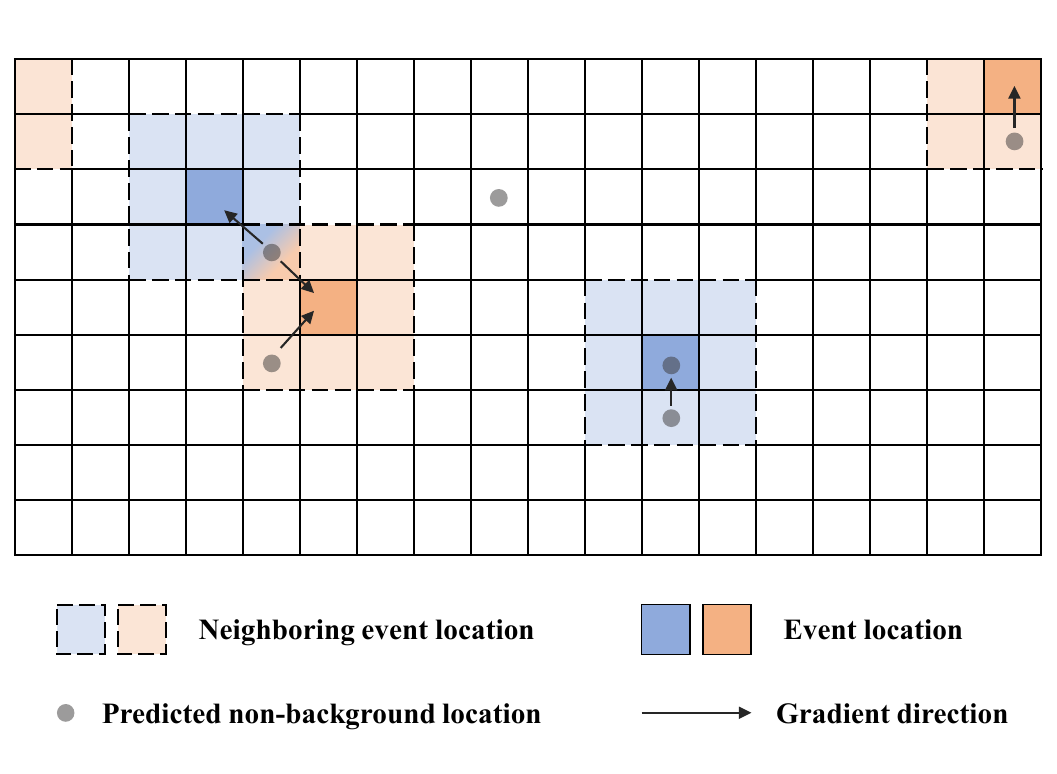}}
\caption{The schematic representation depicts the influence of the asymptotic localization loss function on the model's predictions. The arrows in the diagram indicate the directionality of the guidance provided by the loss function, steering the model towards more accurate predictions.}
\label{cov1}
\end{figure}

\subsubsection{Converging Localization Loss Function}
The converging localization loss $L_{CL}$ helps the model to converge the predicted locations towards the actual non-background area from its surroundings, as shown in Fig.~\ref{cov1}. Initially, the reference target label for each grid cell is transformed by \eqref{target}. 
\begin{eqnarray}
y_{ij}'  = \begin{cases}
   \quad \quad \quad 1 &  \text{if background} \\  
   -N_{bac}/ N_{non\_bac} ,&   \text{if non-background}  
\end{cases}
\label{target}
\end{eqnarray}
where $N_{bac}$ and $N_{non\_bac}$ are the number of background and non-background events, respectively. $y_{ij}'$ represents the transformed reference target of the $i\text{-}$th row and $j\text{-}$th column grid cell.
This transformation is to reduce the influence of the unbalance between background and non-background classes.

After the transformation, the converging localization loss $L_{CL}$ is calculated by \eqref{at}.
\begin{eqnarray}
\begin{aligned}
L_{CL} = \sum_{j=1}^{J} &\sum_{i=1}^{I} (\hat y_{ij}\times y_{ij}^{at})
\end{aligned}
\label{at}
\end{eqnarray}
where $\hat y_{ij}$ represents the predicted probability of a grid cell at the $i\text{-}$th row and $j\text{-}$th column being non-background, and $y_{ij}^{at}$ comprises two elements:
the non-background transformed reference target $y_{ij}'$ and its surroundings' transformed reference target, as defined in \eqref{at1}. 
\begin{eqnarray}
\begin{aligned}
y_{ij}^{at} = y_{ij}'+ AVG(\sum_{j'=j-1}^{(j+1)\%J}\sum_{i'=i-1}^{i+1}y_{i'j'}'-y_{ij}')
\end{aligned}
\label{at1}
\end{eqnarray}
the right side of `+' calculates the mean of the transformed target values of the non-background events' surroundings. 

The $y_{ij}^{at}$ reflects the density of non-background events in a certain area. The greater the density, the smaller $y_{ij}^{at}$. If the model predicts a positive value in a dense area, the resulting $L_{CL}$ will be smaller. By backpropagation, the model predicts a position closer to the dense area, thereby obtaining the correct position prediction.

Ultimately, the three components of the loss function form a union loss function, as presented in \eqref{all}. 
\begin{eqnarray}
\begin{array}{l}
L_{LR}=L_{Class\text{-}MSE}+ L_{AIUR}+ L_{CL} 
\end{array}
\label{all}
\end{eqnarray}

This composite loss function combines each component's strengths to improve the model's performance. $L_{Class\text{-}MSE}$ helps the model predict event classes and take into account location information. $L_{AIUR}$ helps the model align non-background events, making the model pay more attention to non-background events and reducing the impact of extreme class imbalance. $L_{CL}$ focuses on non-background events and their surrounding areas, guiding the prediction results from the surrounding areas to the target location, further reducing extreme category imbalance, and strengthening the model's regression positioning ability.

\subsection{Inference}
After training with SMRL-SELD, during the inference stage, we filter out the non-background grid cells of each frame t. Then convert the position indexes $i$ and $j$ of these grid cells into azimuth and elevation. In this way, all SEDs and DOAs of frame t are obtained. The azimuth and elevation of each grid in 3D space are an angle range rather than a fixed value, so we use the median of the grid cell angle range to represent the angle corresponding to the position of this grid.

\begin{table*}[htbp]
\caption{Comparison of SELD Performance with Different Method on Dev-set-test of STARSS23 and STARSS22 . F$_{20 \degree}$ and LR$_{CD}$ are expressed in percentage.}
\renewcommand\arraystretch{1.1} 
\normalsize
\setlength{\tabcolsep}{3.5pt}
\begin{center}
\resizebox{\linewidth}{!}{
\begin{tabular}{c|ccccc|ccccc}
\hline
\multirow{2}{*}{\textbf{Methods}} & \multicolumn{5}{c|}{\textbf{STARSS23}}                                                                                                                                                      & \multicolumn{5}{c}{\textbf{STARSS22}}                                                                                                                                                        \\ \cline{2-11} 
                                  & \textbf{ER$_{20 \degree} ^{\downarrow}$} & \textbf{F$_{20 \degree} ^{\uparrow}$} & \textbf{LE$_{CD} ^{\downarrow}$} & \textbf{LR$_{CD} ^{\uparrow}$} & \textbf{SELD$_{score}^{\downarrow}$} & \textbf{ER$_{20 \degree} ^{\downarrow}$} & \textbf{F$_{20 \degree} ^{\uparrow}$} & \textbf{LE$_{CD} ^{\downarrow}$} & \textbf{LR$_{CD} ^{\uparrow}$} & \textbf{SELD$_{score} ^{\downarrow}$} \\ \hline

SELDnet\cite{b_single1}                             & 0.570                                    & 29.9                                  & 23.9$\degree$                    & 47.7                           & 0.482                                & 0.654                                    & 27.1                                  & 26.9$\degree$                    & 51.5                           & 0.504                                 \\
ACCDOA \cite{b_ACCDOA}                           & 0.529                                    & 29.4                                  & 23.1$\degree$                    & 49.5                           & 0.467                                & 0.651                                    & 25.6                                  & 24.9$\degree$                    & 51.1                           & 0.506                                 \\
ADPIT  \cite{b_mutil_ACCDOA1}                           & 0.531                                    & 29.9                                  & \textbf{22.1$\degree$}           & 51.2                           & 0.461                                & 0.599                                    & 25.2                                  & 21.3$\degree$                    & \textbf{52.9}                  & 0.484                                 \\
\textbf{SMRL-SELD(Ours)}          & \textbf{0.410}                           & \textbf{38.6}                         & 22.5$\degree$                    & \textbf{59.5}                  & \textbf{0.389}                       & \textbf{0.491}                           & \textbf{36.6}                         & 19.2$\degree$           & 52.1                           & \textbf{0.428}                        \\ \hline
\end{tabular}
}
\end{center}
\label{res}
\end{table*}

\section{Experiment}

\subsection{Experimental Setups}

\subsubsection{Dataset} 
We utilized STARSS23 \cite{b_STARSS23} and STARSS22\cite{b_STARSS22} datasets for training and evaluation. The Sony-TAu Realistic Spatial Soundscapes 2022 (STARSS22) dataset contains multichannel recordings of sound scenes in various rooms and environments and temporal and spatial annotations of prominent events belonging to a set of target classes. STARSS22 includes 121 audio recordings with the duration from 30 seconds to 5 minutes, which are collected in real sound scenes. 
Compared to the STARSS22, STARSS23 maintains all the recordings of STARSS22, while it adds 4 hours of audio distributed between the training and evaluation sets.

\subsubsection{Processing} 
Our model was trained on signals from the FOA array. We performed Short-Time Fourier Transform (STFT) on the 24 kHz audio signals with a hop length of 0.02 seconds and a window size of 0.04 seconds. The resulting spectrograms were then converted to log-mel scales by 64 filter banks. Acoustic intensity vectors \cite{b_iv} were also incorporated into the training process.

\subsubsection{Training} 
We use mixup \cite{b_mixup} and rotation of FOA signals \cite{b_rotation} to augment the training data. The spatial segmentation grid size is set at 10$\degree$, 15$\degree$, and 20$\degree$. Training employed the Adam optimizer, starting with a learning rate of 0.001. Inputs are 5-second signals with a 1-second hop length. 

\subsubsection{Evaluation} 
Four metrics were used for the evaluation\cite{b_single1}: $ER_{20 \degree}$, $F_{20 \degree}$,  $LE_{CD}$, $LR_{CD}$. $ER_{20 \degree}$ and $F_{20 \degree}$ are the location-dependent error rate and F-score, where predictions are considered as true positives only when the distance from the reference is less than 20$\degree$. $LE_{CD}$ is a localization error that indicates the average angular distance between predictions and references of the same class. $LR_{CD}$ is a simple localization recall metric that expresses the true positive rate of how many of these localization predictions are correctly detected in a class out of the total number of class instances. To evaluate the overall performance, we adopted SELD$_{score}$, which is defined as \eqref{at11}.
\begin{eqnarray}
{SELD}_{\text {score }}  \!\!=\! \frac{
[{ER}_{20^{\circ}} \! + \! (1 \! - \! {F}_{20^{\circ}}) \!+ \! \frac{{LE}_{{CD}}}{\pi} +(1 \! - \! {LR}_{{CD}})] 
}{4} \!\!\!\!
\label{at11}
\end{eqnarray}

\subsection{Experimental Results}

\subsubsection{Performance Comparison}

Table~\ref{res} presents the performances of different methods for solving SELD problems. 
We compared our SMRL-SELD with SELDnet\cite{b_single1}, ACCDOA\cite{b_ACCDOA}, and ADPIT \cite{b_mutil_ACCDOA1}.
SELDnet and ACCDOA are single-event detection, whereas ADPIT can handle multiple events, leveraging Permutation-Invariant Training (PIT)\cite{b_pit} to mitigate track permutation issues.
To be fair, our SMRL-SELD and other comparison methods use the same data augmentation strategies. 
All methods achieve better results on STARSS23 than on STARSS22, which may be because STARSS23 has more training data.
Benefiting from using PIT, ADPIT gets a better $LE_{CD}$ on STARSS23 and a better $LR_{CD}$ on STARSS22 than our SMRL-SELD. 
Nevertheless, our SMRL-SELD achieves the lowest SELD$_{score}$ on both datasets, reducing the SELD$_{score}$ by 0.072 on the STARSS23 dataset and 0.056 on the STARSS22 dataset compared to ADPIT. The comparative results prove its effectiveness.

Table~\ref{hard} presents the performance for same-class overlapping occur on different locations, with $\bigtriangleup SELD_{score}$ reflecting changes from Table~\ref{res}. 
On the STARSS22 dataset, all methods, including SMRL-SELD, decline in performance. However, SMRL-SELD has the smallest decline, with just a 0.061 increase in the $SELD_{score}$.
On the STARSS23 dataset, ADPIT and SMRL-SELD performed better. SMRL-SELD does even better than ADPIT, with a 0.027 decrease in the $SELD_{score}$. 
In STARSS23, there are more complex sounds with more than three same-class overlaps, which seems difficult for ADPIT to handle, but SMRL-SELD can handle this complexity better and demonstrated the best performance again.

\begin{table}[htbp]
\caption{SELD Performance Evaluation on Overlapping Events of the Same Class. The Grid Size of the Spatial Segmentation is 15$\degree$.}
\normalsize
\begin{center}
\resizebox{\linewidth}{!}{
\begin{tabular}{c|cc}
\hline
\multirow{2}{*}{\textbf{Method}} & \multicolumn{2}{c}{\textbf{SELD$_{score} ^{\downarrow}$($\bigtriangleup SELD_{score}$)}} \\ \cline{2-3} 
                                 & \textbf{STARSS23}                           & \textbf{STARSS22}                          \\ \hline
SELDnet\cite{b_single1}              & 0.685({\color{red}+0.200})                  & 0.723({\color{red}+0.219})                 \\
ACCDOA \cite{b_ACCDOA}           & 0.651({\color{red}+0.184})                  & 0.717({\color{red}+0.211})                 \\
ADPIT \cite{b_mutil_ACCDOA1}     & 0.443({\color{blue}-0.018})                 & 0.592({\color{red}+0.108})                 \\
\textbf{SMRL-SELD(Ours)}         & 0.362({\color{blue}\textbf{-0.027}})        & 0.489({\color{red}\textbf{+0.061}})        \\ \hline
\end{tabular}
}
\end{center}
\label{hard}
\end{table}

Table~\ref{degree} illustrates the influence of grid cell size on the performance of the SMRL-SELD. It finds that using a 15-degree grid size gives the best results.
The poorer performance at 10$\degree$ and 20$\degree$ could stem from the fuzzy index-to-angle conversion. 
The larger the grid cell, the greater the distance error between the predicted position and the target position. Large grid divisions may also cause non-background events to overlap within the same grid cell.
The smaller the grid, the more detailed the classification needs to be, which increases the difficulty of classification. It is difficult for the model to capture enough information at the existing scale, increasing the possibility of error.

\begin{table}[htbp]
\caption{Comparison of SELD Performance with Different Grid Cell Size. F$_{20 \degree}$ and LR$_{CD}$ Reported in Percentages.}
\renewcommand\arraystretch{1.1} 
\normalsize
\setlength{\tabcolsep}{3.5pt}
\begin{center}
\begin{tabular}{c|cc}
\hline
\multirow{2}{*}{\textbf{Grid Size}} & \multicolumn{2}{c}{\textbf{SELD$_{score}^{\downarrow}$}}   \\ \cline{2-3} 
                                    & \multicolumn{1}{c|}{\textbf{STARSS23}} & \textbf{STARSS22} \\ \hline
10$\degree$                        & \multicolumn{1}{c|}{0.448}             & 0.456             \\
15$\degree$                         & \multicolumn{1}{c|}{\textbf{0.389}}    & \textbf{0.428}    \\
20$\degree$                         & \multicolumn{1}{c|}{0.409}             & 0.461    \\ \hline
\end{tabular}
\end{center}
\label{degree}
\end{table}

\subsubsection{Ablation Study} 
As shown in Table~\ref{ablation}, the ablation study investigates the impact of different parts of the regression localization loss function on the performance of SMRL-SELD.
The first one is only the class mean squared error loss $L_{C\text{-}MSE}$, which increases the SELD$_{score}$ of STARSS23 by 0.052 and the SELD$_{score}$ of STARSS22 by 0.060. The second one is adding the area intersection union ratio loss $L_{AIUR}$, which increases the SELD$_{score}$ of STARSS23 to 0.420, an increase of 0.031, and the SELD$_{score}$ of STARSS22 to 0.452, an increase of 0.024. 
Using the full loss configuration with all components achieves the best performance, producing the lowest SELD$_{score}$.
The results show that all parts of the regression localization loss significantly affected the SMRL-SELD performance.

\begin{table}[htbp]
\caption{Ablation Study Results for SMRL-SELD. The Grid Size of the Spatial Segmentation is 15$\degree$.}
\normalsize
\begin{center}
\resizebox{\linewidth}{!}{
\begin{tabular}{c|cc}
\hline
\multirow{2}{*}{\textbf{Components}} & \multicolumn{2}{c}{\textbf{SELD$_{score} ^{\downarrow}$ ($\bigtriangleup SELD_{score}$)}} \\ \cline{2-3} 
                                     & \textbf{STARSS23}                           & \textbf{STARSS22}                           \\ \hline
$L_{C\text{-}MSE}$                            & 0.441(+0.052)                               & 0.488(+0.060)                               \\
$L_{C\text{-}MSE}+L_{AIUR}$                    & 0.420(+0.031)                               & 0.452(+0.024)                          \\
$L_{C\text{-}MSE}+L_{AIUR}+L_{AL}$             & \textbf{0.389}                              & \textbf{0.428}                              \\ \hline
\end{tabular}
}
\end{center}
\label{ablation}
\end{table}

\section{Conclusions}
We propose Spatial Mapping and Regression Localization for SELD (SMRL-SELD) to enhance the generality in polyphonic environments. SMRL-SELD segments the 3D spatial space, mapping it to a 2D plane, and a new regression localization loss is proposed to help the results converge toward the location of the corresponding event. SMRL-SELD is location-oriented, allowing the model to learn event features based on orientation. Thus, the method enables the model to process polyphonic sounds regardless of the number of overlapping events. We conducted experiments on STARSS23 and STARSS22 datasets and our proposed SMRL-SELD outperforms the existing SELD methods in overall evaluation and polyphony environments.

\bibliographystyle{IEEEbib}
\bibliography{main}

\end{document}